\documentclass[aoas,nameyear,dvips]{arximspdf}
\usepackage{graphics}
\doi{10.1214/09-AOAS261}
\volume{3}
\issue{4}
\pubyear{2009}
\firstpage{1675}
\lastpage{1694}
\begin{document}
\begin{frontmatter}

\title{An integrative analysis of cancer gene expression studies using
    Bayesian latent factor modeling\protect\thanksref{T1}}
\runtitle{Bayesian factor analyses in integrative cancer biology}
\thankstext{T1}{Supported in part by NSF Grant DMS-03-42172 and
National Institutes of Health Grant NCI U54-CA-112952.  Any opinions,
findings and conclusions or recommendations expressed in this work are
those of the authors and do not necessarily reflect the views of the
NSF or NIH.}

\begin{aug}
\author[a]{\fnms{Daniel} \snm{Merl}\ead[label=e1]{dan@stat.duke.edu}\corref{}},
\author[b]{\fnms{Julia Ling-Yu} \snm{Chen}\ead[label=e2]{julia.chen@duke.edu}},
\author[b]{\fnms{Jen-Tsan} \snm{Chi}\ead[label=e3]{jentsan.chi@duke.edu}}
\and
\author[c]{\fnms{Mike} \snm{West}\ead[label=e4]{mw@stat.duke.edu}}

\runauthor{Merl, Chen, Chi and West}
\affiliation{Duke University}

\address[a]{D. Merl\\
Department of Statistical Science\\
Duke University\\
Durham, North Carolina 27708-0251\\
USA\\
\printead{e1}}

\address[b]{J. L.-Y. Chen\\
J.-T. Chi\\
Department of Molecular Genetics\\
\quad  \& Microbiology\\
Institute for Genome Sciences and Policy\\
Duke University Medical Center\\
Box 3382\\
Durham, North Carolina 27710-3382\\
USA\\
\printead{e2}\\
\phantom{E-mail:\ }\printead*{e3}}

\address[c]{M. West\\
Department of Statistical Science\\
Institute for Genome Sciences and Policy\\
Duke University\\
Durham, North Carolina 27708-0251\\
USA\\
\printead{e4}}
\end{aug}

% HISTORY:
\received{\smonth{11} \syear{2008}}
\revised{\smonth{5} \syear{2009}}

% ABSTRACT
\begin{abstract}
We present an applied study in cancer genomics for integrating data
  and inferences from laboratory experiments on cancer cell lines with
  observational data obtained from human breast cancer studies.
  The~biological focus is on improving understanding of transcriptional
  responses of tumors to changes in the pH level of the cellular
  microenvironment.  The statistical focus is on connecting
  experimentally defined biomarkers of such responses to clinical
  outcome in observational studies of breast cancer patients.  Our
  analysis exemplifies a general strategy for accomplishing this kind
  of integration across contexts.  The statistical methodologies
  employed here draw heavily on Bayesian sparse factor models for
  identifying, modularizing and correlating with clinical outcome
  these \textit{signatures} of aggregate changes in gene expression.  By
  projecting patterns of biological response linked to specific
  experimental interventions into observational studies where such
  responses may be evidenced via variation in gene expression across
  samples, we are able to define biomarkers of clinically relevant
  physiological states and outcomes that are rooted in the biology of
  the original experiment.  Through this approach we identify
  microenvironment-related prognostic factors capable of predicting
  long term survival in two independent breast cancer datasets.  These
  results suggest possible directions for future laboratory studies,
  as well as indicate the potential for therapeutic advances though
  targeted disruption of specific pathway components.
\end{abstract}

% KEYWORDS

\end{frontmatter}

%s1 ###
\section{Introduction}
Cancer progression involves a complex interaction of genetic and
genomic factors that jointly subvert normal cell development.  The~genomic component, which encompasses gene expression and regulation,
is substantially impacted by the biochemical composition of the local
environment in which a cell grows. So-called \textit{micro-environmental}
parameters, including levels of oxidation, lactate, acidity,
nutrients of various kinds and other factors affecting physical
interactions between cells, are increasingly studied for their
potential to improve our understanding of cancer biology, and for
their promise to lead to new therapeutic strategies. Changes in such
parameters can impact gene transcription, which in turn impacts
protein production.  Variation in these fundamental parameters can
therefore induce a cascade of effects, producing disruptions of normal
cellular processes in downstream biological pathways [\citet{hanahan}].
For example, changes to the pH level in the cellular environment may
effect glycolysis, thus impacting on numerous genes involved in the
glycolysis pathway.  Some of these genes may also play roles in the
regulation of cell growth, and their suppression may engender
tumorigenesis and promote the aggressive advance of existing cancerous
states.  Microarray gene expression assays can be used to generate
data on the transcriptional response of cancer cells to controlled
manipulations of environmental factors such as pH.  This data is
useful for characterizing these \textit{micro-environmental response pathways}.

Our study concerns changes to cellular pH levels, and the resulting
\textit{neutralization} and \textit{lactic acidosis response pathways}.  Section
\ref{signaturesection} describes the application of sparse Bayesian
regression models [\citet{lucas06};\break \citet{seowest07}] to microarray data
generated through a series of laboratory experiments on cultured
breast tumor cells in which cellular pH levels were manipulated in a
controlled manner.  These analyses yield statistical \textit{expression
  signatures} of the cellular responses to various interventions on
the pH level.  The~main challenge lies in relating these signatures,
and the biological pathways they characterize, to variation in gene
expression across large samples of human breast tumors.  This
integration of in vitro and in vivo data sets is the
driving focus of this and related studies. In addition to comprising a
detailed study of new data and experimental results, through which are
generated several directions for biomedical research, this work
exemplifies an overall strategy for cross-study, integrative analysis
of gene expression data for exploring and relating pathway-related
experimental findings to clinical contexts and patient outcomes.

When considering variability in expression patterns of genes in
observational tumor data, we face questions of differences due to the
differing contexts. It is to be expected that a tumor in vivo
evidences far more complex and heterogeneous biological variation than
in the controlled in vitro setting, and this will be manifest in
measures of gene expression. Normal cell processes held in quiescence
in cell cultures may when active co-regulate the expression of
relevant signature genes in in vivo, confounding the pattern of
expression that was evident in vitro.  Hence, when aiming to
translate experimental findings to tumor populations, thereby
providing a mapping of an in vitro signature to its in vivo counterpart, we require statistical models capable of
discovering and representing the additional complexity surrounding and
interacting with the original response signature.  Section
\ref{statsection} describes our analysis of a large and heterogeneous
breast cancer data set using sparse latent factor models
[\citet{west2003bayes7}; \citet{carvalho2008}] that satisfy these
desiderata. This analysis includes a targeted factor search that
facilitates estimation of statistical factors associated with an
initial set of genes underlying the in vitro experimental
signatures.  The~factors discovered in this way represent a modular
decomposition of the biological patterns evident in the in vivo
breast cancer data, while retaining connections to the experimental
signatures.

Section \ref{biosurvsection} discusses aspects of the biological and
clinical interpretations of these estimated factors, which can be
viewed as a refined in vivo set of summary biomarkers of
variation in the \textit{neutralization} and \textit{lactic acidosis
  response pathways} of these breast cancers. In survival analyses, we
find that these factor model derived biomarkers have substantial
prognostic value in connection with long-term survival and, hence, the
sets of genes comprising these factors warrant further study.  We
present predictive validation of this key finding in analyses of two
separate breast cancer data sets.  We then provide biological
interpretation of one key factor that emerged from the evolutionary
factor search, which plays a key role as a predictive variable in the
survival analyses.  It turns out that this factor is a single component
of a specific biological pathway that has previously been noted as a
risk biomarker in cancer, but not, to date, connected at all into
response pathways linked with variation in cellular pH.  This finding
has generated follow-on biological research and initiated a new line
of experimentation on the role of this pathway in connection with
cancer cell micro-environmental influences.
%Section \ref{conclusionsection} concludes with summary comments.

%s2 ###
\section{Neutralization experiments and analysis}\label{signaturesection}

%s2.1 ###
\subsection{Biological and experimental context}
Investigating the effects of\break changes in the micro-environment in which
cells grow is of increasing interest in cancer research.  The~tumor
micro-environment is typically characterized by oxygen depletion, high
lactate and extracellular acidosis coupled with vascular leakage,
glucose and energy deprivation. These and other micro-environmental
features vary widely across tumors and generally exhibit substantial
temporal and spatial differences in a tumor. Micro-environmental
stresses trigger biochemical changes in cancer cells that directly
modulate physiological, metabolic and ultimately clinical
phenotypes. Improved understanding of the molecular mechanisms of such
tumor responses holds promise for immediate translational impact and
clinical care, as relevant therapies can be brought to bear to modify
the micro-environment.

Currently, with the exception of hypoxia, very little is understood
about how each individual stress affects cellular phenotypes and tumor
progression.  To examine how cancer cells respond to increased acidity
or pH neutralization at different time points, MCF7 cell cultures (a
commonly-used breast tumor cell line) were grown in neutral media and
then exposed to varying interventions in several assays in
parallel. For some cells, lactic acid was added to the medium (25 mM
lactic acid at pH 6.7) for 1 and 4 hours; others cells experienced
strong lactic acidosis conditions (25 nM lactic acid at pH 5.5) for 4
hours.  Similarly, the effects of neutralization were assayed by
shifting the MCF7 cultures from overnight lactic acidosis conditions
at pH 6.7 to neutral regular media at pH 7.4 for 1 and 4 hours.
Control cells were grown in each starting condition (neutral
conditions and lactic acidosis conditions).  The~complete set of
experiments is summarized in Table \ref{exptable}.  The~mRNA extracted
from each of the resulting $n=27$ batches of MCF7 cultures was
purified using Ambion miRVana RNA purification kits and standard
microarray assays were performed using Affymetrix U133 Plus 2 Genechip
platforms. All raw microarray data were preprocessed using RMA
[\citet{rma}], the log (base 2) scale output of which were used in all ensuing
statistical analyses.

%t1 ###
\begin{table}%[t]
\caption{Summary of neutralization/acidosis experiments.
Cell entries indicate the number of replicates per experimental group}\label{exptable}
\begin{tabular*}{\textwidth}{@{\extracolsep{\fill}}lccccc@{}}
 \hline
  & \multicolumn{5}{c@{}}{\textbf{Exposure condition}}\\[-6pt]
 & \multicolumn{5}{c@{}}{\hrulefill}\\
 & \multicolumn{2}{c}{\textbf{pH 7.4}}& \multicolumn{2}{c}{\textbf{pH 6.7}} & \multicolumn{1}{c@{}}{\textbf{pH 5.5}} \\[-6pt]
 & \multicolumn{2}{c}{\hrulefill}& \multicolumn{2}{c}{\hrulefill} & \multicolumn{1}{c@{}}{\hrulefill} \\
 \textbf{Growth condition} & \textbf{1hr} & \textbf{4hr} & \textbf{1hr} & \textbf{4hr} & \textbf{4hr}\\
\hline
pH 7.4 & 3$\times $ & 3$\times $ & 3$\times $ & 3$\times $ & 3$\times $\\
pH 6.7 & 3$\times $ & 3$\times $ & 3$\times $ & 3$\times $ &\\
\hline
\end{tabular*}
\end{table}

%s2.2 ###
\subsection{Cellular response signatures}\label{sparsereg}
Quantitative summaries of the cellular responses to lactic acidosis
and neutralization treatments were obtained using a standard sparse
multivariate regression model [\citet{lucas06}; \citet{seowest07}].  We analyzed
$19{,}375$ genes (technically, probe-sets from the Affymetrix array; we
will use ``gene'' and ``probe'' interchangeably) whose median
expression level is at least 5.5 and whose expression ranges more than
$0.5$-fold across the $n=27$ experimental samples.  Let $X^{\mathrm{exp}}$
denote the $19{,}375 \times 27$ matrix of expression values.  Rows
represent genes and columns correspond to three replicate samples for
each of the following experimental groups: (i) control (pH~$7.4\rightarrow 7.4$) at 1~hour; (ii) control at 4 hours; (iii)
lactic acidosis (pH $7.4\rightarrow 6.7$) at 1~hour;  (iv) lactic acidosis at 4 hours; (v) neutralization
(pH~$6.7\rightarrow 7.4$) at 1~hour; (vi) neutralization at 4 hours;
(vii) acidic growth (constant pH of 6.7) at 1~hour; (viii)
acidic growth at 4 hours; (ix) strong lactic acidosis (pH
$7.4\rightarrow 5.5$) at 4 hours.  Let $H^{\mathrm{exp}}$ denote the $11 \times
27$ design matrix where the first 8 rows contain binary indicators for
effects associated with differential expression relative to the 1hr
control group: 1hr lactic acidosis effect, 1hr neutralization effect,
1hr acidic growth effect, 4hr control effect, 4hr lactic acidosis
effect (relative to 4hr control), 4hr neutralization (relative to 4hr
control), 4hr acidic growth effect (relative to 4hr control) and 4hr
strong lactic acidosis effect (relative to 4hr control).  The~last
three rows contain artefact control factors derived from the first
three principle components of the expression levels associated with
the AFFX series control genes included on the Affymetrix microarrays.
These control genes are not variably expressed in humans, and so
patterns of variation across samples manifest in control genes
represents systematic errors arising from different experimental
conditions.  Use of these artefact control factors provides
opportunity for sample-specific correction of artefactual effects on
genes that may otherwise result in false-discovery or obscure
meaningful biological variation
[following \citet{lucas06} and \citet{carvalho2008}].  After deriving the
artefact control factors, rows corresponding to Affymetrix control
genes are removed from subsequent analyses.

The~model for the
expression of gene $g$ in sample $i$ is
\[
x_{g,i}^{\mathrm{exp}}  =  \mu_g + \sum_{k=1}^{11}\beta_{g,k}h_{k,i}^{\mathrm{exp}} +
\nu_{g,i}
\]
or in matrix from
\[
X^{\mathrm{exp}}  =  \mu 1 + BH + N,
\]
where  $\mu_g$ denotes the mean expression of
gene $g$ in the 1hr control samples,
each $\beta_{g,k}$ is the change in
expression of gene $g$ due to design factor  $k$, and the
$\nu_{g,i}$ are independent,  normally distributed idiosyncratic noise terms representing
residual biological variation, experimental and measurement errors with individual
variances $\psi_g.$
Sparsity is induced via prior distributions that place
positive probability on $\beta_{g,k}=0$ for each $g,k$ pair, and
resulting posterior analysis allows investigation of posterior
sparsity patterns via probabilities  $\pi_{g,k}^* = \operatorname{Pr}(\beta_{g,k}\ne 0|X^{\mathrm{exp}}).$
Full details follow \citet{lucas06} and prior specifications, including priors
for the~$\mu_k,$ variance parameters and all hyper-parameters, are given in
 \ref{appendix}.  Posterior inference via MCMC is achieved using the BFRM software [\citet{bfrm}].

\begin{figure}%[htbp]

\includegraphics{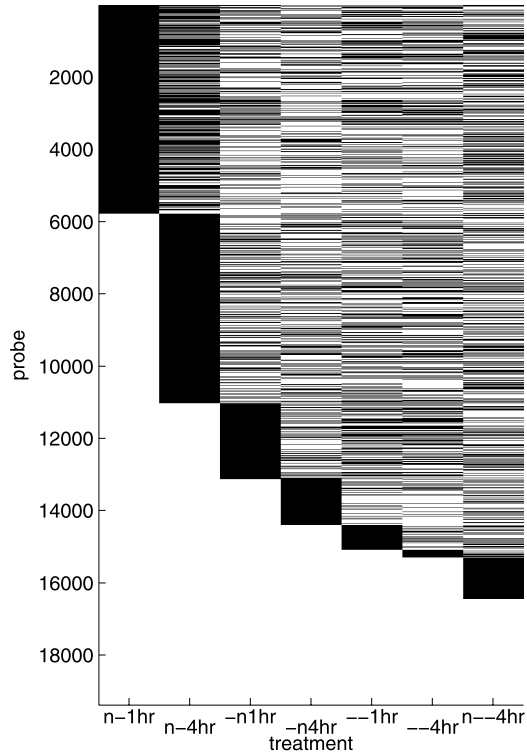}

     \caption{Neutralization signature skeleton: black indicates genes $g$ (rows)
    with posterior probability $\pi_{g,k}^*>0.99$ for each experimental
    group $k$ (columns). Genes are ordered to emphasize which genes are unique
    to each successive experiment relative to the previous.}\label{sigskeleton}
\end{figure}

Figure \ref{sigskeleton} broadly illustrates genes uniquely associated
with individual treatment effects as well as those involved in
multiple responses.  This gives some indication of the degree of
intersection of the cellular pathways being queried by the different
treatments.  Across the 8 treatments, the sparsity, as measured by the
percent of genes for which $\pi_{g,k}^*>0.99$, ranges from $29\%$ (4~hour neutralization) to $46\%$ (4~hour lactic acidosis).  The~fold-change associated with the involved genes ($2^{|\beta_{g,k}|}$
for $g$ such that $\pi_{g,k}^*>0.99$) ranges from 1.06$\times $ to 13$\times $, with a
mean of 1.4$\times $.

%f1 ###

%f2 ###
\begin{figure}%[htbp]

\includegraphics{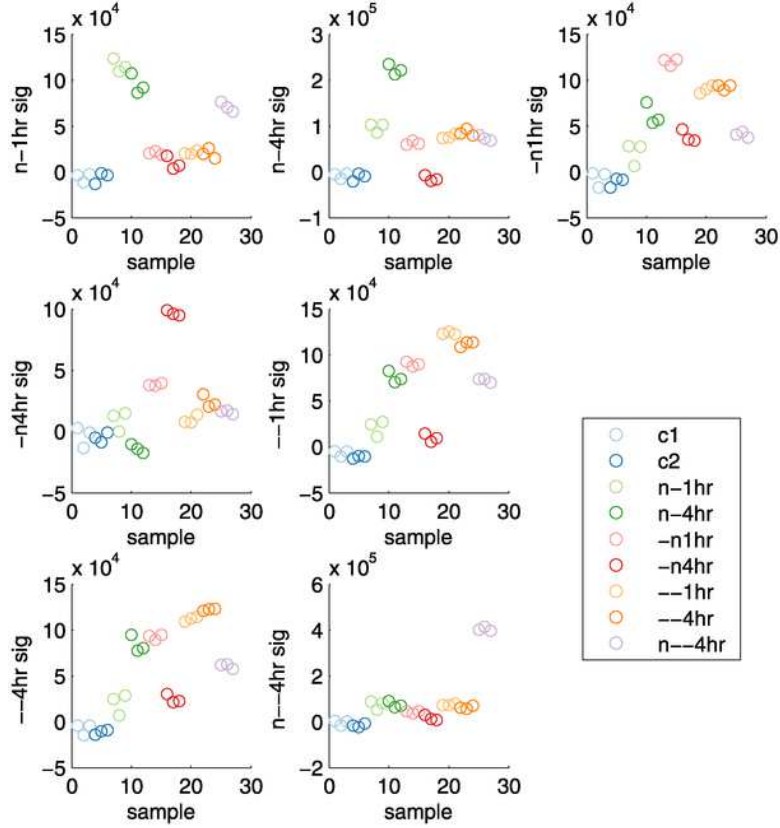}

  \caption{Neutralization treatment
        signature scores ($s_{k,i}$) for each sample in the original
        study. Separate treatment groups are color coded.} \label{sigcohorts}
\end{figure}

The~cellular response to each treatment, also called the \textit{signature} of the treatment, is characterized by estimated effects
$\beta_{g,k}^* = E(\beta_{g,k} | \beta_{g,k}\ne 0,X^{\mathrm{exp}})$ together
with the $\pi_{g,k}^*.$ The~ability of each signature to uniquely
identify the treatment it reflects can be further explored using
summary \textit{signature scores} as defined in \citet{lucas09}.  Based on
posterior means $\beta_{g,k}^*$ and $\psi_g^*,$ let
\[
s_{k,i} = \sum_{g=1}^{19{,}375}
\beta_{g,k}^*x_{g,i}^{\mathrm{exp}}/{\psi_g^*}
\]
define the score for treatment signature $k$ on sample $i.$ This
expression is derived from the data-driven component of the Bayes
factor that weighs the evidence in favor of the given signature
describing the variation in a sample ($p(x_i | h_{k,i}=1)/p(x_i |
h_{k,i}=0)$).  Figure \ref{sigcohorts} shows the values of the scores
associated with 7 treatment signatures plotted across samples.  As
expected, the highest scoring samples for each signature are those
upon which that signature is based, but important connections between
signatures can be identified on the basis of other high- or
low-scoring treatment groups.  For example, there is a inverse
relationship between the 4hr acidosis score and the 4hr neutralization
score.  Also evident is the similarity between the 1hr and 4hr acidic
growth signatures, which can also be inferred through the large
intersection of the genes defining the two signatures (Figure
\ref{sigskeleton}).

%s3 ###
\section{Latent factor analysis of breast tumor gene expression}\label{statsection}

%s3.1 ###
\subsection{In vivo  breast cancer data}\label{millersection}
The~primary goals of this study are to uncover shared structures in
the cell response signatures defined above, and to quantify the extent
to which these structures can be used to predict clinical phenotypes
in real human cancers.  Here we make use of the gene expression data
for a collection of 251 surgically removed breast tumors as reported
in \citet{miller}.  Affymetrix 133A and 133B
GeneChip microarrays were generated for each tumor sample, and
relevant clinico-pathological variables were collected for each
patient.  This included age at diagnosis, tumor size, lymph node
status (an indicator of metastatic cancer) and Elston histological
grade (a categorical rating of malignancy as deemed by pathologists).
Molecular assays to identify the presence of absence of mutations in
the estrogen receptor (ER), progesterone receptor (PgR) and P53 genes
were also performed. These data are representative of a variety of
different presentations of human breast cancer on these clinical
measures.

\begin{figure}%[htbp]

\includegraphics{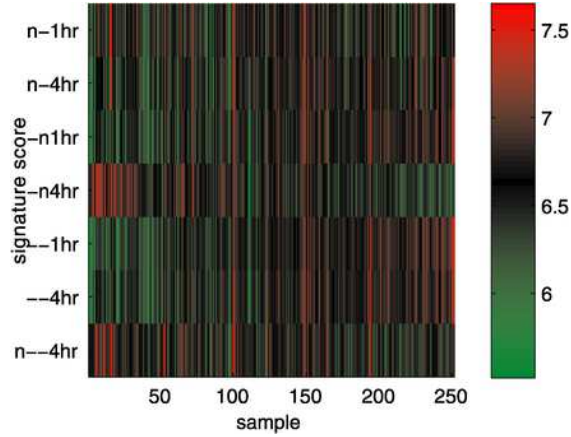}

     \caption{Initial evaluation of neutralization signature levels
      across tumor samples.  Samples are ordered by first principle
      component to emphasize dominant signature
      gradients.}\label{sigscores}
      \end{figure}

We first evaluate the signature score as defined above for each tumor.  These
scores are then standardized across samples so that each vector of 27
scores for a particular signature has mean and variance equal to the
mean gene-specific expression and mean gene-specific variance.  This
transformation places the signature scores on the same scale as gene
expression in the tumor data set, thus enabling a ``metagene''
interpretation of a vector of scores [\citet{west01}; \citet{pittman04}]; see
Figure \ref{sigscores}.
The~relationships between the tumor signature scores bear some
similarities to those observed in the cell line study.  There is once
again evidence of correlation between the two acidic growth signatures
and the 1hr neutralization signature.  These three signatures, in
turn, display patterns opposite that of the 4hr neutralization
signature.  The~patterns are less prominent, however, than was evident
in the cell culture data.  Although the variation in these scores
presumably relates, in part, to underlying biological variation in the
activity of the lactic acidosis and neutralization response pathways
within these tumors, as mentioned above, the set of genes
characterizing the in vivo effects of lactic acidosis and
neutralization may differ substantially from those characterizing the
in vitro responses as a result of the more complex interactions
with other cellular processes.

%f3 ###

We thus aim to refine our evaluation of the response pathway activity
levels in the tumor data by using the signature scores as initial
``anchors'' in an analysis using sparse latent factor models.  The~main idea is to define statistical factors on sets of genes related to
these initial scores, and to link in other genes that may connect with
the different response pathways active in vivo.  This is
accomplished as follows.

%s3.2 ###
\subsection{Sparse factor model specification}\label{sparsefac}
Sparse latent factor models represent common patterns in gene
expression via latent factors in which the factor-gene relationships
are sparse; this notion of statistical sparsity\break is key for
representing the intersecting subsets of genes potentially\break related to
underlying networks of biological pathways
[\citet{west2003bayes7};\break \citet{seowest07}; \citet{lucas08};\break \citet{carvalho2008}].  The~form of
the statistical model is an extension of the sparse regression
model. A key part of our analysis strategy stems from augmenting the
$44{,}592 \times 251$ matrix of gene expression data for the tumor data
with the 7 values of the projected treatment signature scores.  Let
$p=44{,}592+7=44{,}599$ and $n=251$, and let $X^{\mathrm{obs}}$ denote the $p
\times n$ matrix in which the first 7 rows are the projected scores
across tumor samples, and rows $8-p$ are the gene expression values.
Here we will make use of $K=4$ artefact control factors derived from
the first four principal components of the control genes of the breast
tumor microarrays.  A latent factor model consisting of $L$ latent
factors is therefore
\[
x_{g,i}^{\mathrm{obs}}  =  \mu_g + \sum_{k=1}^K \alpha_{g,k}\lambda_{k,i} + \sum_{l=K+1}^{K+L}\alpha_{g,l}\lambda_{l,i} + \nu_{g,i}
\]
or, in matrix form,
\[
X^{\mathrm{obs}}  =  \mu 1 + A \Lambda + N,
\]
where: (i) the first $K$ rows of the $(K+L) \times n$ matrix
$\Lambda$ are the known artefact controls; (ii) the remaining
$L$ rows contain latent factor scores; (iii) the first $K$
columns of the $p\times (K+L)$ matrix $A$ are regression parameters on
the artefact controls (changing notation from the earlier $\beta$ to
$\alpha$ for notational convenience here); (iv) the remaining
$L$ columns of $A$ are factor loadings parameters relating factors to
genes and to the projected scores; and (v) $A$ is sparse, with
sparsity pattern to be inferred along with estimation of nonzero
values.  The~model is completed by assigning sparsity priors over
columns of $A,$ precisely as was done for $B$ in the sparse regression
model; prior specification for $A,$ variance components and other
hyper-parameters follows default recommendations for the BFRM
framework (see \ref{appendix}).

Flexibility in representing potentially complicated patterns
underlying expression is achieved using nonparametric Bayesian
Dirichlet process models for the factor scores.  The~$L$-vectors $
(\lambda_{K+1,i},\ldots,\lambda_{K+L,i})'$, representing the latent
factor values on tumor sample $i$, are modeled as draws from an
unknown latent factor distribution subject to a Dirichlet process
prior with a multivariate normal base measure.  This standard
nonparametric mixture model allows great flexibility in adapting to
nonnormal structures commonly manifest in factor scores
[\citet{carvalho2008}; \citet{bfrm}].

Ensuring the identifiability of latent factors requires the use of a
modified prior on $A$ such that the leading L rows have an upper
triangle of zeros and positive upper diagonal elements; that is, for
$g=1\dvtx L,$ we have $\alpha_{g,g+K}>0$ and $\alpha_{g,l}=0$ for $l>g+K$.
The~first $L$ variables in $X^{\mathrm{obs}}$ then represent ``founders''
of the $L$ latent factors, with variable $g$ associated with a
$\alpha_{g,g}$-fold change in expression due to factor $g,$
$(g=1,\ldots,L).$ It also defines an hierarchical dependence on the
factors, namely,
\begin{eqnarray*}
x_{1,i}^{\mathrm{obs}} & = & \cdots + \alpha_{1,K+1}\lambda_{K+1,i} + \nu_{1,i}, \nonumber\\
x_{2,i}^{\mathrm{obs}} & = & \cdots + \alpha_{2,K+1}\lambda_{K+1,i} + \alpha_{2,K+2}\lambda_{K+2,i}+\nu_{2,i},\nonumber
\label{facreg}\\
x_{3,i}^{\mathrm{obs}} & = & \cdots + \alpha_{3,K+1}\lambda_{K+1,i} + \alpha_{3,K+2}\lambda_{K+2,i}+\alpha_{3,K+3}\lambda_{K+3,i}+\nu_{3,i}\nonumber
\end{eqnarray*}
and so on.  This structure aids the interpretation of the latent
factor loadings as representing interconnected components of a complex
biological process.  The~latent factor scores $\lambda_{i,l}$ quantify
variation across tumors for these expanding levels of complexity, with
each additional factor accounting for variation in observed gene
expression unaccounted for by the previous set of factors.  With our
use of projected in vitro signature scores here as the first 7
variables, the first 7 factors will now represent patterns underlying
co-variation in expression of sets of genes that link indirectly to
these treatment signatures. Additional factors then reflect other
dimensions of common variation in the set of genes analyzed.

%s3.3 ###
\subsection{Targeted factor search}\label{facmodsearch}
Decomposition of the patterns of variation evident in the tumor gene
expression data into latent factors proceeds through evolutionary
model search, full details of which appear in
\citet{carvalho2008} and \citet{bfrm}.  The~evolutionary model search provides a
computationally efficient approximation to the computationally
prohibitive full factor analysis on the entire set of genes, and
produces full posterior results for the final set of factors and
genes.  A key novelty of this approach is that we exploit the
sensitivity of the model search procedure to its initial configuration
in order to explore the space of factor models surrounding an initial
model containing $7$ latent factors and representing only the 7
response metagenes.  By construction, these initial factors are each
defined, or ``founded,'' by the neutralization/lactic acidosis
treatment scores, thereby ensuring that the model search is primarily
concerned with patterns of variation related to these particular
response pathways.

Evolution of this initial model proceeds as follows.  Samples from the
joint posterior distribution of model parameters are obtained through
MCMC.  Based on these fitted values, we impute inferences for all
genes $g>7$ that are not currently included in the model, as described
in \citet{carvalho2008}.  Thus, after fitting the initial factor model
which considers only the signature scores, we examine expression
levels of the full set of $44{,}000+$ genes for evidence of association
with the current factors.  The~imputation process generates
approximate probabilities $\pi_{g,l}^* = \operatorname{Pr}(\alpha_{g,l}\ne 0
| X^{\mathrm{obs}})$ for all such genes $g.$ Genes are ranked on the basis of
these probabilities, and the model is then expanded to include a small
number of the genes with largest values of the projected
$\pi_{g,l}^*.$ The~model is then refitted to this expanded sample, and
if appropriate, the number of factors is increased in order to adapt
to additional common patterns of expression variation now evident in
the increased set of variables being modeled.  This process is
repeated until no new genes or factors can be added, or until the
model reaches a designated maximum size.  More details on the search
strategy, including control parameters governing model expansion, are
given in \ref{appendix}.

%f4 ###
\begin{figure}%[htbp]

\includegraphics{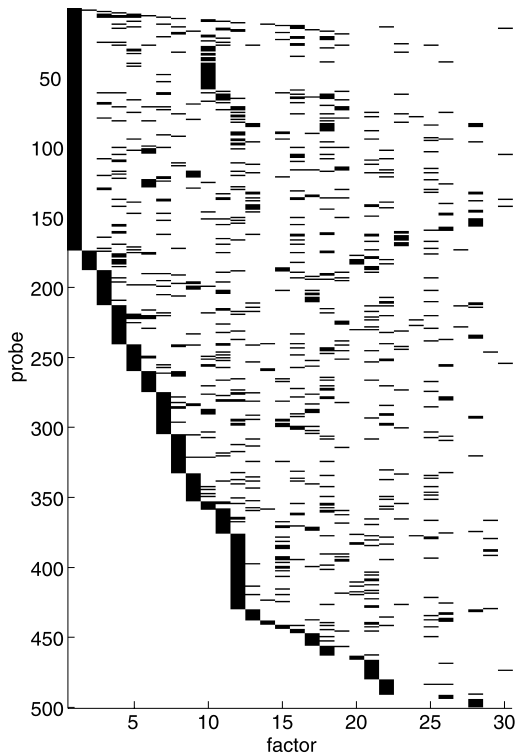}

     \caption{Skeleton of fitted factor loadings for  tumor data.  Black indicates
      variable-factor loadings with $\pi_{g,l}^*>0.99.$ The~first 7 variables are the
       projected neutralization scores, followed by 493 genes
       reordered for a clear visual presentation of the sparsity
       structure of, and cross-talk in, gene-factor loadings.}\label{facskel}
\end{figure}

The~initial 7-gene, 7-factor model evolved under this process to reach
a terminal size of 500 variables (the designated maximum)
incorporating 30 latent factors.  Figure \ref{facskel} shows the
skeleton of the factor structure, in terms of major patterns of
gene-factor relationships.  The~ordering of the factors is determined
by the model search procedure, and represents the incremental
improvement to model fit provided by each subsequent factor.  In this
sense, each subsequent factor builds upon the complexity modeled by
the previous factors.  The~leading 7 factors
correspond to the following signatures, respectively: 4hr lactic
acidosis, 1hr lactic acidosis, 1hr neutralization, 4hr strong lactic
acidosis, 4hr acidic growth, 1hr acidic growth, and 4hr
neutralization.  Like their in vitro signature counterparts, the
in vivo factors loadings contain a great deal of sparsity.  Of the 493
genes included in the final model, only 333 are among those identified
in the in vitro signature analysis.  Factor 1, founded by the
the 4hr lactic acidosis signature score, has 173 genes with nonzero
loadings at the $0.99$ probability threshold, compared to $8909$ in
the in vitro signature.
% This reduced gene set still provides a
% good model for the 4hr lactic acidosis signature, as can be seen by
% comparing the factor model fitted values for the signature scores
% against the original scores (Figure \ref{fittedsigs}).

%f5 ###
\begin{figure}%[htbp]

\includegraphics{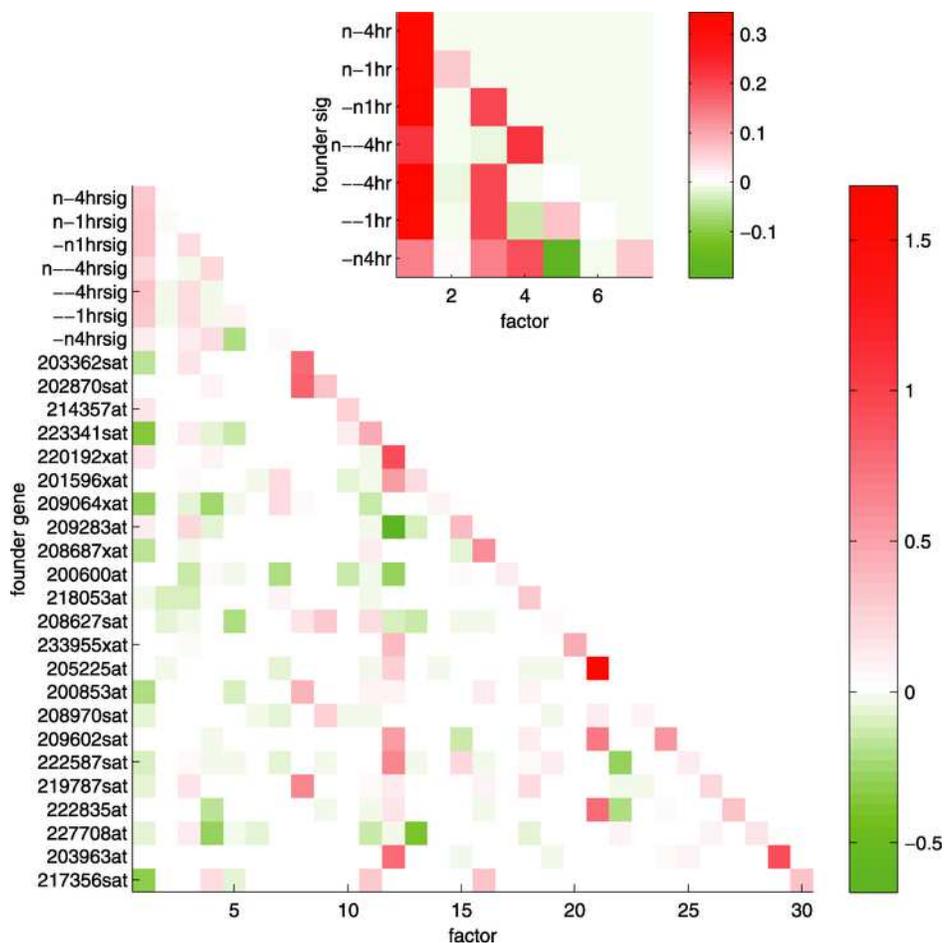}

 \caption{Heat map showing the magnitudes of the fitted factor loadings $\alpha_{g,l}^*$
for the first $L=30$ rows of $A$. The~founder gene for each factor is designated by its
U133$+$ probe ID.  The~terms in each row determine a linear combination
of latent factors that predict the observed expression levels of the founder gene. }\label{crosstalk}
  \end{figure}

Posterior estimates of the factor loadings ($\alpha_{g,l}^* =
E(\alpha_{g,l}|\alpha_{g,l}\ne 0,X^{\mathrm{obs}})$) aid in generating further
insights. In particular, the upper portion of the estimated loadings
matrix sheds light on the structure of connections between latent
factors; see Figure \ref{crosstalk}.  As described in Section
\ref{sparsefac}, one interpretation of a row $A$ is as a set of
coefficients determining a linear combination of factor scores that
predict the gene expression vector for the corresponding variable.
The~inset of Figure \ref{crosstalk} shows that the fitted values of
all 7 of signature scores involve positive contributions from factor
1, the factor version of the 4hr lactic acidosis signature.  Thus, the
pattern of 4hr lactic acidosis signature activity across samples
describes a fundamental pattern of pathway activation that underlies
the activity patterns of the other 6 signatures.  The~seventh factor
(i.e., the factor representation of the 4hr neutralization signature)
sits atop this hierarchy of pathway complexity, represented as a
linear combination of factors 1 (4hr neutralization), 3 (1hr
neutralization), 4 (4hr strong lactic acidosis) and 5 (4hr acidic
growth), plus the additional pattern of expression unique to this
pathway.

% \begin{figure}[htbp]
% \begin{center}
%   \includegraphics[width=5in]{figs/fittedsigs.png}
%    \begin{minipage}{.85\textwidth}
%    \caption{Comparison of factor model-based fitted values with original
%     signature scores.  The~fitted scores, even though modelled by a
%     significantly reduced set of genes as compared to
%     the original signatures, demonstrate very high similarity to the original
%     signature scores.}\label{fittedsigs}
%     \end{minipage}
% \end{center}\end{figure}

%s4 ###
\section{Biomedical connections of factor profiles}\label{biosurvsection}

%s4.1 ###
\subsection{Factor-based prediction  of long term survival}\label{survivalsection}
The~in vivo latent factors linked to neutralization pathways
represent complexity in the patterns of expression, and therefore in
the levels of underlying biological pathway activation evident across
the tumor samples.  For this reason, latent factors can be regarded as
candidate biomarkers of physiological states that link to these
pathways. Our study explores this using the posterior mean factor
scores $\lambda_{l,i}^*$ as candidate predictors in a survival analysis
of the breast cancer patient data.

% edited to here
We use Weibull regression models of patient survival that draw on the
30 estimated neutralization/lactic acidosis pathway factors, the 7
original projected signature scores and the clinical covariates
available for this data set [\citet{miller}].  The~latter include
histologic grade, ER mutation status, node status, P53 mutation
status, PgR mutation status, tumor size and age at diagnosis.  This
analysis allows both integration and comparison of the prognostic
value of these traditional markers with specific pathway-related
signature scores, and their latent factor representations---an
integrative clinico-genomic analysis.  Let $t_i$ denote the survival
time of patient $i.$ The~Weibull density function is $p(t_i|a,\gamma)
= at_i^{a-1}\exp(\eta_i - t_i^ae^{\eta_i}),$ where $a>0$ is the index
parameter and $\eta_i = \gamma'y_i$ the linear predictor based on a
covariate vector $y_i.$ We explore subsets of covariates and
regression model uncertainty using Bayesian shotgun stochastic search
[\citet{hans2007}; \citet{sss}]. This generates a list of regression covariate
subsets and the corresponding posterior regression model probabilities
for use in Bayesian model averaging for survival prediction and in
exploring relevance of variables. Details of model and prior
specification follow defaults in the SSS software [\citet{sss}] as noted
in \ref{appendix}.

\begin{figure}[b]

\includegraphics{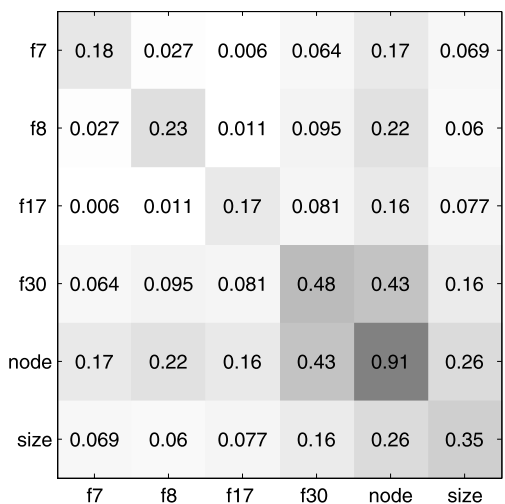}

     \caption{Pairwise  inclusion probabilities for the top
      6 predictor variables in breast cancer survival analysis.
      Darker colored tiles indicate higher
      probabilities.}\label{pairprobs}
  \end{figure}

Figure \ref{incprobs} shows posterior probabilities for each of the
46 candidate covariates.  Nodal status emerges as the leading
predictor of long term survival, followed by latent factor 30 and then
tumor size.  Note that none of the original signature scores, and no
other clinical variables, receive appreciable probability.  That nodal
status provides the best predictor of survival is to be expected.
Previous studies [\citet{west01}; \citet{pittman04}; \citet{dressman06}] have shown that
nodal status is not well predicted by gene expression and that
combined use of nodal status with gene expression predictors can
improve survival prognosis. Hence, it seems that the
information content of the nodal status predictor is unlikely to
overlap with that of any factor score.  This can also be clearly seen
through the pairwise inclusion probabilities in Figure
\ref{pairprobs}.

%f6 ###
\begin{figure}[t]

\includegraphics{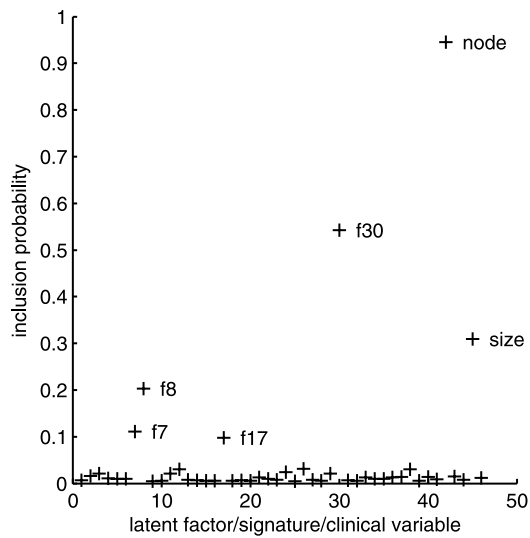}

     \caption{Posterior inclusion probabilities for the 46 candidate covariates in the Weibull
      models. The~candidate covariates include the 30 estimated
      latent factors, followed by the 7 original signature scores,
      followed by 10 traditional clinical covariates
         (Elston grades 1, 2 and 3, ER, node
      status, P53, PgR status, tumor size, age at diagnosis).}\label{incprobs}
  \end{figure}

%f7 ###

The~pairwise inclusion probability of factor 30 and nodal status is
close to the marginal probability of
factor 30; however, the pairwise inclusion probability of factor 30
with any of the other factors is far less than any of the marginal
inclusions probabilities of those factors; thus, factor 30 is clearly a dominant and
preferred expression-based biomarker of survival risk.

%f8 ###
\begin{figure}[t]

\includegraphics{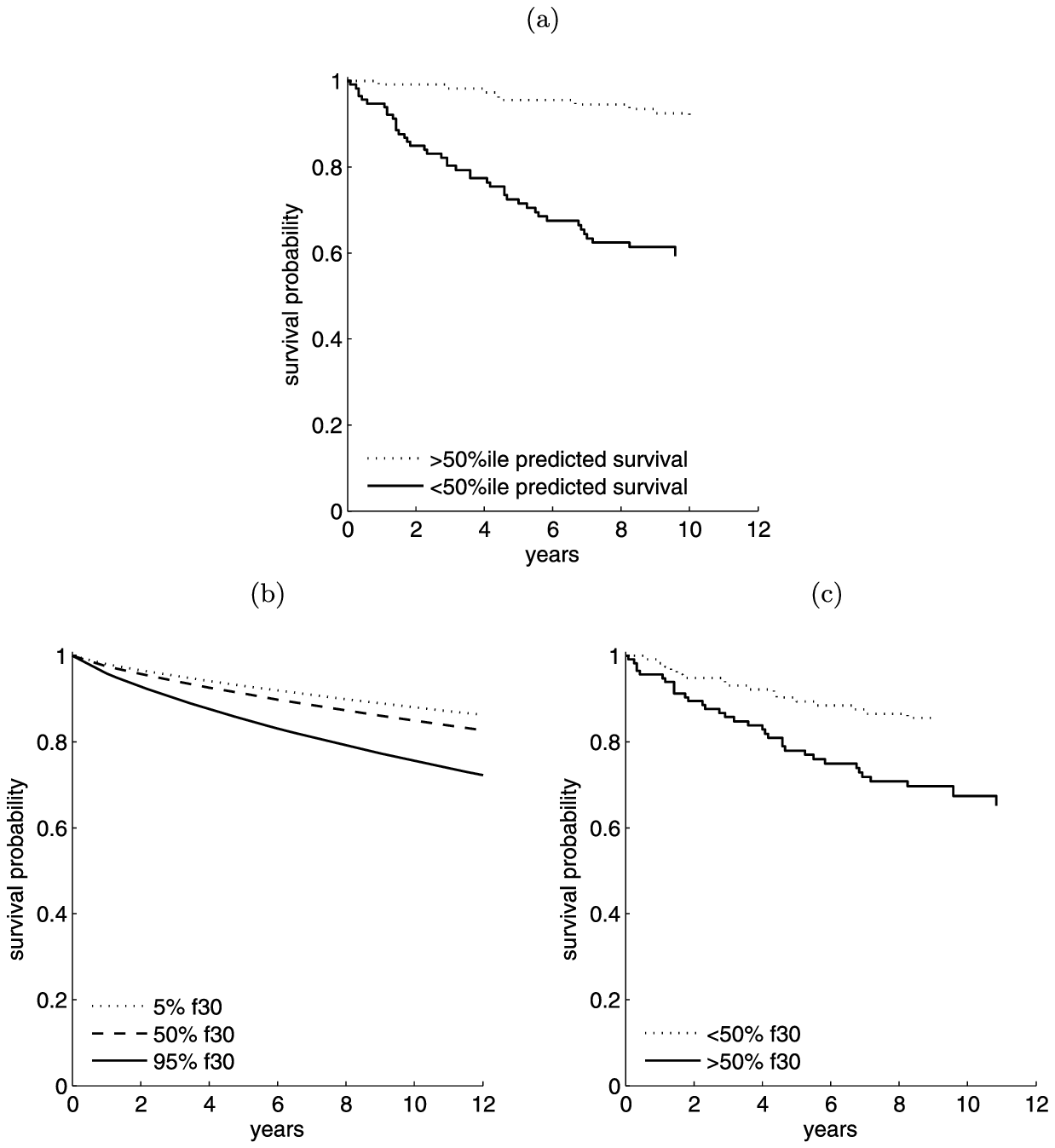}

     \caption{\textup{(a)} Kaplan--Meier curves demonstrating stratification
     of Miller data into high- and low- risk groups, based on the fitted Weibull mixture.
     \textup{(b)} Estimated survival curve associated with varying levels of
     factor 30, holding all other predictors at their median values.
     \textup{(c)} Kaplan--Meier curves demonstrating stratification of
     Miller data into high- and low- risk groups, based solely on the value of factor 30.}\label{survplots}
     \end{figure}

Posterior and predictive inferences are formally based on a mixture of
1000 Weibull survival models, mixed with respect to their posterior
probabilities.  For each patient $i,$ we can compute the implied
predicted survival function at her covariate vector $y_i$ and identify
the predicted median survival $m_i;$ this is the predicted median
survival time for a \textit{future} patient who has the same covariates
as patient $i.$ Figure \ref{survplots}(a) shows a Kaplan--Meier
survival plot of the Miller et al. data simply stratified by
$m_i\le m$ or $m_i>m,$ where $m = \operatorname{median}\{ m_i,
i=1,\ldots,n\}.$ There is approximately a 30\% difference in the
empirical 10-year survival probability between patients cohorts
stratified crudely on this basis, as a simple visual of the relevance
of the included covariates.  By way of focusing on factor 30, we plot
the model-averaged survival curve for a hypothetical patient whose
covariates are held constant at their median values in the data set,
save for variation in the factor~30~score; see Figure
\ref{survplots}(b), where factor 30 is set at its $5$th, $50$th
and $95$th percentiles in the data set, all other covariates
remaining fixed.  The~estimated effect of variation in factor 30 alone
accounts for approximately 20\% of the difference in 10-year patient
survival between the high-risk and low-risk subgroups.  This
prediction is confirmed by considering the Kaplan--Meier curves formed
by stratifying the patients on the basis of the patient-specific
factor 30 value compared to the median across samples; see Figure
\ref{survplots}(c).  The~pattern of gene expression comprising
the loading associated with factor 30 warrants further investigation,
to which we will return in Section \ref{biof30}.

%f9 ###
\begin{figure}[b]

\includegraphics{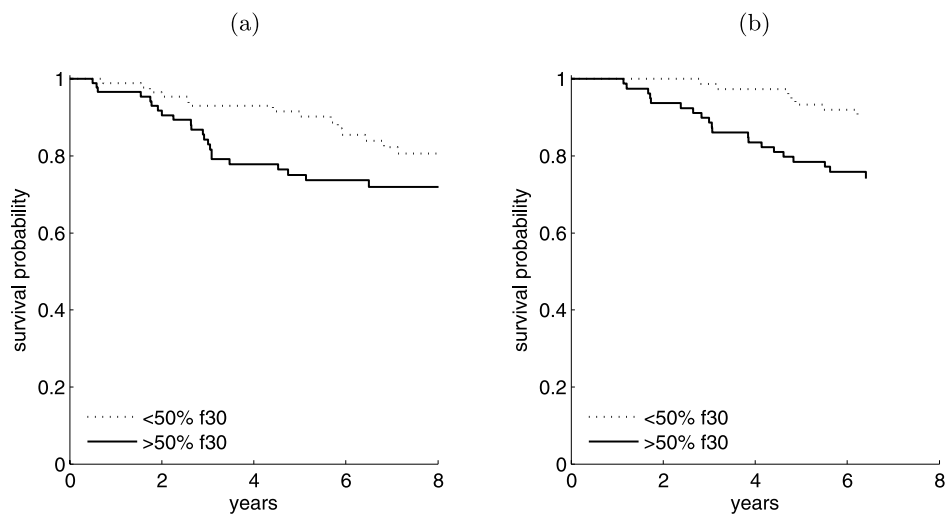}

 \caption{Kaplan--Meier curves demonstrating stratification of
\textup{(a)} the \protect\citet{pawitan05} patient samples, and
\textup{(b)} the \protect\citet{sotiriou06} patient samples (right)
  into high- and low- risk groups based on imputed values of factor
  30, as identified in the latent factor analysis of the Miller data.}
  \label{bcsurvpredict}
  \end{figure}

%s4.2 ###
\subsection{Out-of-sample factor projection}\label{facproj}
It is critical to evaluate whether or not the above results can be
confirmed through out-of-sample prediction.  We do this with two
additional breast cancer data sets: that of \citet{pawitan05}, consisting of 159 primary breast tumors assayed on
Affymetrix U133A and U133B chips, and that of \citet{sotiriou06}, consisting of 189 primary breast tumors assayed on
U133A chips.

Fixing all factor model parameters at their posterior means, we can
directly predict values of the latent factors for each new patient;
see\break \ref{appendix} and [\citet{lucas08}].  Note that this calculation is purely
predictive; no model fitting nor additional analysis of the two
validation data sets was performed.  Using the predicted latent factor
vectors, we can produce the same survival plots for these data,
stratifying each of the two new patient cohorts on the basis of their
factor 30 scores as above; see Figure \ref{bcsurvpredict}, as compared
to Figure \ref{survplots}(c).
The~association between low factor 30 scores and good prognosis
remains evident in these out-of-sample predictions that draw on
different patient populations.  Further, the differences between high
and low risk groups is comparable across all three samples.

\begin{figure}[t]

\includegraphics{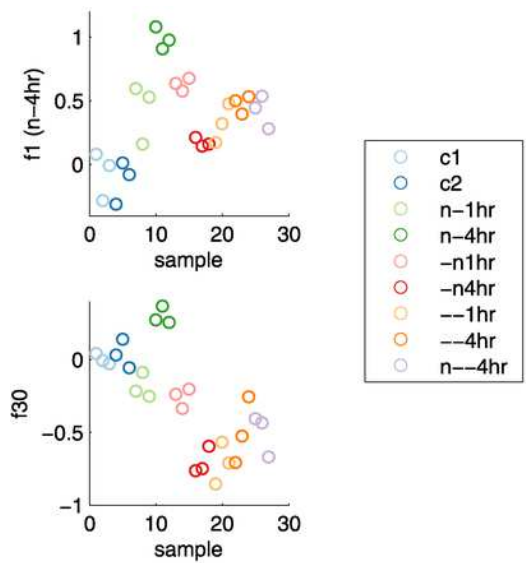}

\caption{Predicted values of factors 1 and 30 for each sample in the original experimental study.
Separate treatment groups are color coded.}\label{f1f30reverse}
\end{figure}

The~robustness of factor 30 as a prognostic biomarker provides strong
support for the view that it reflects a biologically meaningful module
of gene expression.  By evaluating the predicted factor scores in the
original experimental data, we are able to establish that factor 30,
despite its relatively late incorporation to the model, is linked to
the 4-hour lactic acidosis pathway.  Figure \ref{f1f30reverse}
compares the predicted factor scores for factors 1 and 30 as evaluated
in the experimental data.  Factor 1, which is founded by the 4-hour
lactic acidosis signature, is in fact comparable to the original
signature score as depicted in Figure \ref{sigcohorts}.  Factor 30 has
its highest values in the original 4-hour lactic acidosis samples, but
shows a different pattern of activity across the other sample cohorts.
In particular, factor 30 appears to be repressed in the samples
associated with the 4-hour neutralization and acidic growth
treatments.  This implies that factor 30 may characterize some
critical intersection between these pathways that is itself related to
tumor aggressiveness.

%f10 ###

\eject
%s4.3 ###
\subsection{Biological evaluation of prognostic factor 30}\label{biof30}

Having established the clinical relevance of factor 30, the task
remains to ascribe to it biological meaning. The~loading of factor 30
is comprised of only 6 gene probe sets for which
$\pi_{g,l}^*>0.99$. Four of these, including the founder gene, are
related to the phosphoglycerate kinase 1 (PGK1) gene, while the other
two are related to a neuronal cell death-related protein and the CEGP1
protein.  The~factor is characterized by overexpression
($\beta_{g,k}>0$) of the PGK1 and neuronal cell death proteins and
suppression ($\beta_{g,k}<0$) of CEGP1.

A literature search generates detailed biological information on PGK1,
and its role in the glycolysis pathway where it is fundamental to
cell\break
growth and metabolism.  PGK1 catalyzes the reversible conversion
of\break
1,3-diphosphoglycerate to 3-phosphoglycerate with the generation of
one molecule of ATP and this represents an important step in
glycolysis pathways. In addition, PGK1 has been reported to induce
other processes related to cancer progression, such as conferring a
multi-drug resistant (MDR) phenotype [\citet{duan}] and affecting tumor
angiogenesis through affecting secreted plasmin [\citet{lay}]. Previous
studies have also shown that elevated levels of PGK1 predict \textit{poor} survival outcomes in lung cancers [\citet{chen03}], and that
PGK1 can often be expressed at high levels in pancreatic [\citet{hwang}]
and renal [\citet{unwin}] cancers.  The~association between high factor
30 levels and poor prognosis indicates a similar relationship between
PGK1 and survival may exist for breast cancers.

Since PGK1 is an important component of glycolysis pathways, our
findings here may implicate glycolysis activities in poor patient
survival. This is supported by previous findings that expression of
glycolysis pathways and PGK1 are repressed by lactic acidosis
[\citet{chen08}]. Factor 30 links the neutralization pathway response
signatures to a clear \textit{PGK1 factor} that may now serve as a
biomarker of one key aspect of tumor responses to changes in pH with
the potential to aid in predicting follow-on changes in tumor
metabolism via glycolysis pathway activation.  Further evaluation of
this chain of relationships is now initiated and will be explored
using independent methods such as tumor tissue microarrays
[\citet{chen03}]. Since PGK1 and glycolysis pathways are also controlled
by hypoxia [\citet{chi}], these results also highlight their potential
roles as integral mediators of multiple micro-environmental factors
affecting tumor progression and clinical outcomes.

\section*{Acknowledgments}
The~authors are grateful to an editor of AOAS and an anonymous
reviewer for constructive comments on this manuscript.

\eject
\begin{supplement}[id=supp:code+etc]
\sname{Supplement A}
\stitle{Software and Data}
\slink[doi]{10.1214/09-AOAS261SUPPA}
\sdatatype{url}
\slink[url]{http://ftp.stat.duke.edu/WorkingPapers/08-34.html}
\sdescription{
  This site contains all materials needed to reproduce the reported
  analyses.  This includes all data files, control files for the BFRM
  and SSS software, and MATLAB functions for producing graphical summaries.}
\end{supplement}

\begin{supplement}[id=appendix]
\sname{Supplement B}
\stitle{Appendix}
\slink[doi]{10.1214/09-AOAS261SUPPB}
\slink[url]{http://lib.stat.cmu.edu/aoas/261/supplement.pdf}
\sdatatype{.pdf}
\sdescription{The appendix Merl et al. (\citeyear{merl09}) provides further details
on prior specifications in the sparse regression and sparse latent factor models.
The~appendix also contains details on the control parameters for the evolutionary
factor search and shotgun stochastic search, and describes the procedure for imputing factor scores in new samples.\looseness=-1}
\end{supplement}
\iffalse

\fi

\printaddresses

\end{document}